\documentclass[%
 reprint,
superscriptaddress,amsmath,amssymb,aps,prb,]{revtex4-2}
\usepackage[dvipsnames]{xcolor}
\usepackage{graphicx}
\usepackage{bm}

\begin{document}

\preprint{APS/123-QED}

\title{Plasmons in a layered strange metal using the gauge-gravity duality\\}

\author{S.T. Van den Eede}
\affiliation{
Institute for Theoretical Physics and Center for Extreme Matter and Emergent Phenomena, Utrecht University, Princetonplein 5, 3584 CC Utrecht, The Netherlands}

\author{T.J.N. van Stralen}
\affiliation{
Department of Applied Physics, Eindhoven University of Technology, Groene Loper 19, 5612 AP Eindhoven, The Netherlands \\
}
\author{C.F.J. Flipse}
\affiliation{
Department of Applied Physics, Eindhoven University of Technology, Groene Loper 19, 5612 AP Eindhoven, The Netherlands \\
}
\author{H.T.C. Stoof}

\affiliation{
Institute for Theoretical Physics and Center for Extreme Matter and Emergent Phenomena, Utrecht University, Princetonplein 5, 3584 CC Utrecht, The Netherlands}
\date{October 2023}

\begin{abstract}
In an attempt to understand the density-density response of the cuprate superconductors, we study plasmons in a layered strange metal using the Gubser-Rocha model. The latter is a well-known bottom-up holographic model for a strange metal that is used here to describe the strongly repulsive on-site interactions between the electrons in each copper-oxide (CuO$_2$) layer, whereas the long-range Coulomb interactions are incorporated by a so-called double-trace deformation. To be able to model the bilayer cuprates more realistically, we consider in particular the case of two closely-spaced CuO$_2$ layers per unit cell. In the response we then obtain for vanishing out-of-plane momentum both an optical and an acoustic plasmon, whereas for nonvanishing out-of-plane momentum there are two acoustic plasmon modes. We present the full density-density spectral functions with parameters typical for cuprates and discuss both the dispersion and the lifetime of these plasmon excitations. Moreover, we compute the conductivity after introducing disorder into the system. Finally, we also compute the loss function to facilitate a comparison with experimental results from electron energy loss spectroscopy.

\end{abstract}

\maketitle

\section{\label{sec:level1}Introduction}
Strange metals have puzzled physicists for decades, as certain defining properties of these metals cannot be easily understood by standard condensed-matter physics. Most importantly, the electrical resistivity is perfectly linear in temperature \cite{proust2019,cooper2009,bruin2013}, up to the melting point of the material, and it exceeds the Mott-Ioffe-Regel limit \cite{hussey2004}. These peculiar properties are presumed to be linked to the unusually strong interactions between the charge carriers in a strange metal \cite{anderson1997,plakida2010}. In this paper we focus on layered strange metals, in particular the cuprate superconductors, which is a class of materials characterized by copper-oxide (CuO$_2$) layers stacked on top of each other \cite{park1995}, with insulating charge reservoirs in between them whose composition can be altered to dope the CuO$_2$ layers. 
In the normal phase cuprates exhibit strange-metal behavior within a limited doping and temperature range \cite{zaanen-sc2015}.
Moreover, cuprates are superconductors up to relatively high temperatures and critical temperatures of up to $135$ K have been observed \cite{schilling1993}.
It is unknown why the critical temperature is so high in these materials, but it is suspected that a thorough understanding of the strange-metal phase is required to answer this question \cite{zaanen-sc2015}.
Studying the behavior of cuprate superconductors above their critical temperature, where they become strange metals, might therefore help to increase the critical temperature further towards room temperature in the future. Achieving room-temperature superconductivity at atmospheric pressure is one of the greatest goals of condensed-matter physics. 
In this paper, we use a holographic model to investigate the properties of charge-density oscillations, better known as plasmons, in the strange-metal phase of cuprate superconductors.
The plasmons in this class of materials have been studied in several experiments before \cite{nucker1989,bozovic1990,nucker1991,levallois2016,mitrano2018,nag2020,hepting2022,hepting2023,bejas2023} but so far there has not been a single theoretical framework which describes every aspect of these materials.

It is plausible that an improved understanding of these experiments requires a more sophisticated theory which models the effects of the strong on-site Coulomb (Hubbard-U) repulsion between the electrons in the strange-metal phase. One way to achieve this is to apply a technique that originates from string theory, known as the gauge-gravity duality or AdS/CFT correspondence \cite{maldacena1999}. This correspondence conjectures that there is a relationship between a bulk gravity theory in an anti-de Sitter (AdS) spacetime and a conformal field theory (CFT) on its boundary. Since this is a bulk-boundary correspondence it is also known as the holographic principle, which has proven in recent years to effectively describe low-energy properties of strongly interacting systems \cite{hartnoll2016}. The specific model we use in this paper is the Gubser-Rocha model \cite{gubser2010}, a special case of an Einstein-Maxwell-dilaton model. In this model the entropy scales linearly with temperature, and hence the resistivity is also linear in temperature, which is one of the defining features of a strange metal.
Moreover, recent ARPES experiments have been accurately described by the Gubser-Rocha model \cite{smit2021}. 

Ultimately, and most importantly for our purposes, this model gives a long-wavelength density-density response function for a single CuO$_2$ layer. This response function can in principle only be obtained numerically, but to obtain more analytical insight we use a very accurate hydrodynamic approximation instead. The Gubser-Rocha response function describes strongly interacting but `neutral' electrons, implying that it describes the strong on-site interactions inside each layer, but not yet the long-range effects of the Coulomb interactions which are crucial for the existence of plasmons and thus need to be incorporated separately. 
To incorporate the long-range Coulomb interactions in the framework of the gauge-gravity duality, we need to perform a so-called double-trace deformation of the conformal field theory \cite{witten2002}. This has been done in a layered geometry before \cite{mauri2019} using a different neutral response function that follows from the Einstein-Maxwell gravity theory without the dilaton, but that model does not lead to the description of a strange metal. So, our objective is to determine, using this same double-trace deformation, whether the Gubser-Rocha holographic model for a strange metal can lead to an improved understanding and interpretation of experimental results.

Since our approach consists of a bottom-up holographic computation, we need to realize that response functions are only determined up to an overall constant. 
The reason for this is that Newton's constant $G$ in the gravity theory is not determined from our condensed-matter system.
We can remedy this issue by noting that the density-density response function of the layered strange metal ultimately contains a plasma frequency, which we can use to fix this single undetermined constant such that the holographically obtained plasma frequency matches exactly the experimentally measured plasma frequency.
Throughout this paper, we show how this can be achieved very explicitly, after which we present our final results in the form of the density-density spectral function that depends solely on material parameters that can in principle be determined experimentally. The spectral function will most clearly display the plasmon modes of the charge-density fluctuations, with a sharp peak denoting a long-lived collective mode since the width of the peak determines its lifetime. Moreover, it is directly related to the energy-loss function, which has been observed experimentally.\\

\begin{figure}[h!]
        \centering
        \includegraphics[width=0.3\textwidth]{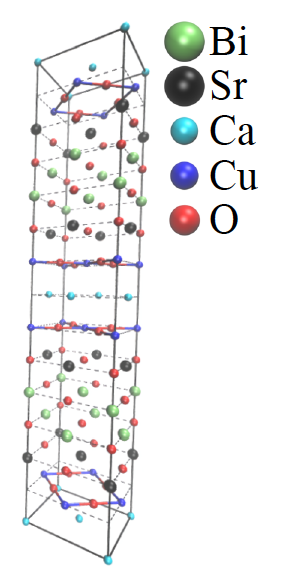}
        \caption{The unit cell of Bi-2212 \cite{thijs2023}. Adjacent pairs of CuO$_2$ planes are rotated by 45 degrees.}
        \label{fig:bi2212}
\end{figure}

Although our approach turns out to be very general and almost completely analytical, we consider in this paper for concreteness the cuprate Bi$_2$Sr$_2$CaCu$_2$O$_{8+x}$ (Bi-2212) which is in the class of bismuth-based cuprates. A schematic representation of the unit cell is displayed in Fig. \ref{fig:bi2212}, which shows that the crystal structure of this particular cuprate is slightly more complex than the most basic layered structure. Namely, it contains pairs of closely-spaced CuO$_2$ planes, and these pairs are in turn separated by a larger distance. The unit cell contains four, instead of two, CuO$_2$ planes because the adjacent pairs of CuO$_2$ planes are rotated by 45 degrees with respect to each other.
However, in our model we do not consider this rotation, since it does not affect the long-wavelength physics on the scale of many unit cells in the in-plane direction. Thus we effectively construct a model with only two CuO$_2$ layers per unit cell. We focus on this particular cuprate because we study this material in our experimental research group in Eindhoven using electron energy loss spectroscopy (EELS) \cite{thijs2023}.

The structure of the paper is as follows. We start with briefly presenting the holographic theory to describe the two-dimensional CuO$_2$ planes. We give the gravitational action for the Gubser-Rocha model and we further explain why we choose this particular model. We also discuss the hydrodynamic approximation of the Gubser-Rocha density-density response function in two dimensions and its corresponding spectral function. Then we introduce for a single layer the long-range Coulomb interactions by performing a double-trace deformation on the single-layer result and we show how this leads to a plasmon mode with a square-root dispersion in the density-density spectral function. From here, we first add another identical layer at a nonzero distance from the first one and derive the density-density spectral function for this bilayer system. Thereafter, we periodically stack these pairs of layers on top of each other to form an infinite crystal which resembles Bi-2212. We again determine and discuss the density-density spectral function in this case. As mentioned above, the plasma frequency we obtain is used to fix the single unknown holographic parameter. Furthermore, we also consider the special limit in which each CuO$_2$ layer is separated by the same distance, to verify that our band structure for two layers per unit cell in that limit exactly reduces to the result for a single layer per unit cell. Next, we compute the conductivity in the case of Planckian dissipation in the system. Finally, we construct the EELS loss function to be able to compare our findings with experiments.
\\

\section{\label{sec:level1} Holographic theory of a two-dimensional strange metal}
In this section we go over some of the mathematical details to describe the strong interactions in the two-dimensional CuO$_2$ planes. We give the gravity action used and the resulting density-density response function. For the rest of the paper we use the expression obtained here. If desired it is possible to use also a different description of the strong in-plane interactions, but the addition of long-range Coulomb interactions will proceed in exactly the same way as presented after this brief summary of holography.

As mentioned previously, we here use the holographic principle \cite{maldacena1999} to derive the single-layer  response. This principle states that a strongly interacting quantum field theory is equivalent or dual to a classical gravitational theory with one additional spatial dimension. This is also known as the AdS/CFT correspondence. The anti-de Sitter spacetime is the curved bulk spacetime with the conformal field theory on its boundary that is located at $r \rightarrow \infty$, where $r$ is the additional space coordinate of the bulk spacetime. More specifically, we use a version of the Einstein-Maxwell-dilaton model proposed by Gubser and Rocha \cite{gubser2010}. This model is dual to a quantum field theory characterized by `semi-local' quantum-critical behavior. This implies that the only momentum dependence in the electron self-energy is in the exponent, i.e., $\hbar \Sigma(\omega,\textbf{k}) \propto \omega(-\omega^2)^{\nu_k-1/2}$. It represents a quantum-critical theory because the correlation length diverges and its dynamical exponent obeys $z = \infty$. The significance of this result is that it agrees with experimental observations. For example, upon tuning the adjustable parameters in the holographic model such that $\nu_{k_F} \equiv \alpha$, this self-energy can reproduce the `power-law liquid' model, $\hbar \Sigma''(\omega,\textbf{k}) \propto \omega^{2\alpha}$, which very accurately describes the experimentally observed electron self-energy in Angle-Resolved Photo-Emission Spectroscopy (ARPES) measurements near the Fermi surface in the nodal direction \cite{reber2019}. There is even another, more recent, ARPES experiment that confirms the momentum dependence in the exponent and shows that it can accurately describe the deviations from the `power-law liquid' model away from the Fermi surface \cite{smit2021}. These experiments thus indicate that the Gubser-Rocha model describes some aspects of the strange-metal phase, although there are other properties of the strange metal that might not yet be accurately described by this model. For example, the anomalous scaling of the Hall angle \cite{blake2015,amoretti2016,ahn2023}. Although this quantity is not relevant for this paper as we consider no external magnetic field, we are aware that there might be need for a more advanced model which could describe all of these properties simultaneously. 

Next, we give the action and explain how this action describes certain properties of a strange metal. The gravitational action for the model is \cite{mauri2022}

\begin{eqnarray} \label{EMDaction1}
    && S_{GR} = S_{ct} + \frac{c^4}{16\pi G} \int dr dt d^2x\sqrt{-g} \\
    && \times \left[ R - \frac{(\partial_\mu \phi)^2}{2} + \frac{6}{L^2}\cosh{\left( \frac{\phi}{\sqrt{3}}\right)} - \frac{e^{\phi/\sqrt{3}}}{4 g_F^2} F_{\mu\nu} F^{\mu\nu} \right] \nonumber ,  
\end{eqnarray}\\
where $r$ is the additional spatial dimension of the bulk spacetime, $g$ is the determinant of the metric tensor, $R$ is the Ricci scalar and $\phi$ is the dimensionless scalar field known as the dilaton \cite{charmousis2010,gouteraux2011}. Moreover, $F_{\mu\nu}$ is the electromagnetic field strength tensor. Its coupling constant is 
$g_F^2 = {c^4\tilde{\mu}_0}/{16\pi G}$,
with $\tilde{\mu}_0$ the dimension of a magnetic permittivity m\! kg\! C$^{-2}$. Then, $L$ is the anti-de Sitter radius, which is the radius of curvature of the AdS spacetime. Finally, $S_{ct}$ contains the boundary counterterms that ensure that we have a well-defined boundary problem and that the theory is properly renormalized.
The dilaton field $\phi$ is responsible for being able to describe the typical strange-metal behavior of linear-in-T resistivity and gives also a linear momentum dependence in the exponent $\nu_k$ of the correlations. 
We can rewrite the action into dimensionless quantities, by defining lengths in terms of $L$ and energies in terms of $\hbar c/L$, and then the prefactor of the action becomes $N_G \equiv c^3L^2/16\pi\hbar G$, which is related to the large-N number of species of the boundary QFT \cite{zaanen2015}. We come back to this in principle unknown constant and explain how to fix it by looking at the experimentally observed plasma frequency.
The thermodynamics of the two-dimensional strange metal is described by a background solution to the following equations: the Einstein field equations for the metric $g_{\mu\nu}$, the Maxwell equations for the $U(1)$ gauge field $A_\mu$ and the Klein-Gordon equation for the dilaton field $\phi$. These equations can be obtained by varying the above action.

For the background the solutions to the equations of motion are function of $r$ only. We then have a set of equations ($g_{tt}=-1/g_{rr}$, $g_{xx}=g_{yy}$, $A_t$, $\phi$), which supports a fully analytical black-hole solution with non-zero temperature and entropy \cite{gubser2010}.
To compute the response functions, however, we also need to consider small external perturbations of this background and we have to linearize the gravitational equations around the analytical black-hole solution. Then we obtain a coupled set of equations for the fluctuations ($\delta g_{tt}$, $\delta g_{tx}$, $\delta g_{xx}$, $\delta g_{yy}$, $\delta A_t$, $\delta A_x$,  $\delta \phi$) that can only be solved numerically. According to the holographic dictionary, finding a solution to the linearized equations of motion with infalling-wave boundary conditions at the black-hole horizon allows us to extract all the retarded Green's function of the system, and thus also the desired density-density response function $\Pi(\omega, \textbf{q})$, by studying the near-boundary behavior of the field fluctuations \cite{zaanen2015}.

In this manner we arrive at the objective of this section, the two-dimensional single-layer response describing the strong short-range interactions. In the low-temperature regime and at energies and momenta much smaller than the Fermi energy and Fermi momentum, respectively, we can use a hydrodynamic approximation to obtain \cite{kovtun2012}

\begin{equation} \label{strongresponse}
    \Pi(\omega,q) = \frac{q^2(\omega \mathcal{D} +i v_s^2 D_d \chi q^2)}{\omega^3 +i \omega^2 q^2(2D_s+D_d)-\omega q^2 v_s^2-i v_s^2 D_d q^4},
\end{equation}
where we have used rotational invariance to write $q \equiv |{\bf q}|$ as a scalar. In addition, $\mathcal{D}$ is the Drude weight, $\chi$ is the hydrodynamic compressibility, and we also define two diffusion constants $D_s$ and $D_d$ that correspond to sound diffusion and charge diffusion, respectively. Finally, $v_s$ is the speed of sound in the material.
\begin{figure}
        \centering
        \includegraphics[width=0.45\textwidth]{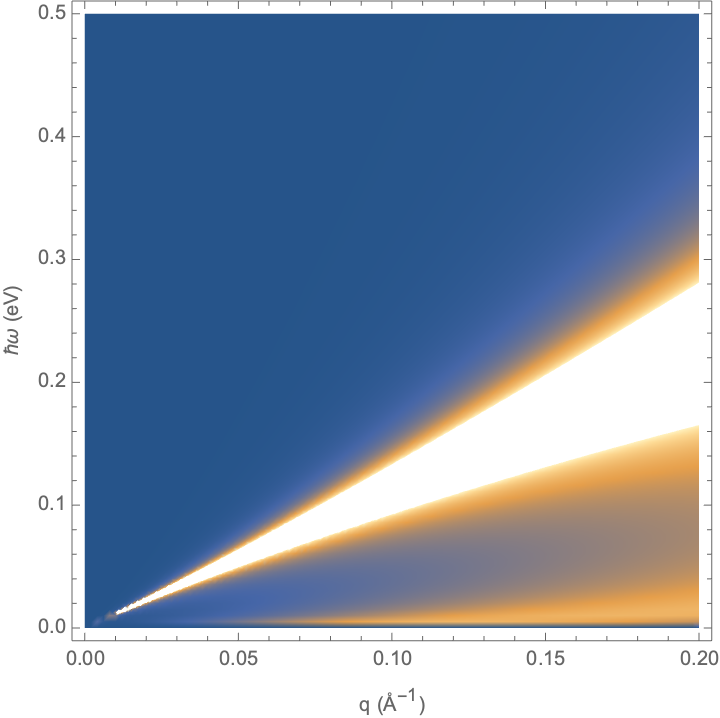}
        \caption{Density-density spectral function $-\Pi''$ of a single layer with only strong short-range interactions. This plot contains a linear sound mode, $\omega=v_s q$, and a diffusive mode, $\omega=-i D_d q^2$. The temperature is fixed at room temperature $T=293$ K.}
         \label{fig:singlelayerstrong0.5}
\end{figure}

\begin{figure}
     \centering
        \includegraphics[width=0.46\textwidth]{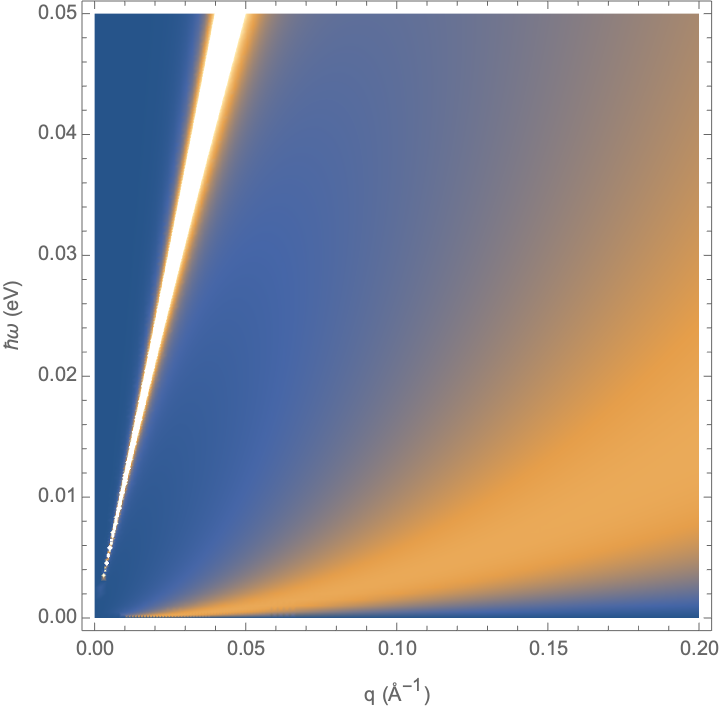}
        \caption{Density-density spectral function as in Fig. 2. In this case we have $\hbar \omega \leq 0.05$ eV, showing more clearly the diffusive mode. }
        \label{fig:diffusionstrong}
    \label{fig:singlelayerstrong}
\end{figure}

In Figs. \ref{fig:singlelayerstrong0.5} and \ref{fig:singlelayerstrong} we show the density-density spectral function $-\Pi'' \equiv -$Im$\, \Pi$. We use values for the variables in Eq. (\ref{strongresponse}) typical for cuprates. We elaborate on these variables in the section 'Derivation of parameters', below. In a spectral function the intensity of modes is plotted as a function of momentum and frequency or energy, thus clearly showing the associated dispersion. It also captures the broadening of the modes, which tell us about the lifetime of the mode. If the width is small the mode has a long lifetime. In Figs. \ref{fig:singlelayerstrong0.5} and \ref{fig:singlelayerstrong}  there is a linear sound mode, $\omega=v_s q$, instead of a typical plasmon mode expected in the presence of long-range Coulomb interactions and screening and with a square-root dispersion $\omega \propto \sqrt{q}$. The speed of sound is $v_s = 0.76 v_F \simeq 1.14$ eV \AA $/ \hbar = 1.73 \times 10^5$ m\! s$^{-1}$. This value of the Fermi velocity of Bi-2212 is chosen because this specific cuprate is studied by our group in Eindhoven \cite{thijs2023}. In the limit $T=0$ the diffusion constants vanish, as they show the same linear behavior in temperature as the resistivity, and Eq. (\ref{strongresponse}) simplifies to  
\begin{equation} \label{strongt=0}
    \Pi(\omega,q) = \frac{q^2\mathcal{D}}{\omega^2-v_s^2q^2},
\end{equation}
which indeed contains the sound mode $\omega=v_s q$. We use this equation in later sections to derive the plasmon dispersion when we have introduced the long-range Coulomb interactions. At non-zero temperatures the density-density response function contains also a diffusive mode with $\omega = -i D_d q^2 + \mathcal{O}(q^4)$, which can be seen more clearly in Fig. \ref{fig:diffusionstrong}. We analyze the response function for $\omega \ll q$, which means we neglect in the denominator the cubed and squared terms in $\omega$. Then multiplying both nominator and denominator with $\omega-i D_d q^2$, so as to take most easily the imaginary part, this leaves us with the following formula for the spectral function
\begin{equation} 
    -\Pi''(\omega,q) =\omega~ \frac{\chi-\mathcal{D}/v_s^2}{D_dq^2}, \, \, \text{for} \, \omega \ll q.
\end{equation}
So for small $\omega$ the intensity approaches zero linearly at a fixed value of $q$, confirming what is shown in Fig. \ref{fig:singlelayerstrong}. Furthermore, we see in Fig. \ref{strongresponse} that for larger momenta the diffusive mode and the sound mode merge together. \\

\section{\label{sec:level1} Plasmon modes }
\subsection{\label{sec:level2} Single-layer plasmons}
Now, having an appropriate response function incorporating the strong but short-range interactions in the two-dimensional strange-metal layer, long-range Coulomb interactions are introduced. We do this by coupling dynamical photons to the density current $J^\mu$. Thus in the language of string theory we perform a so-called double-trace deformation \cite{witten2002,mueck2002} of the conformal field theory. In references \cite{mauri2019,zaanen2019} it is explained more physically how to achieve this, but in practice this means adding a boundary term to the gravitational action from Eq. (\ref{EMDaction1}) leading ultimately to 
\begin{eqnarray} \label{actionEMD3}
    S &=& \frac{1}{2}\int dt d^2{x} dz \int dt' d^2{x}' dz' \nonumber\\
    && \times J^\mu (\textbf{x},z,t) \Pi^{-1}_{\mu\nu}(\textbf{x},t;\textbf{x}',t') J^\nu (\textbf{x}',z',t')  \nonumber\\
    && -\int dt d^2{x} dz \left( \frac{1}{4} \epsilon F_{\mu \nu}F^{\mu \nu} - e A_\mu J^\mu  \right),
\end{eqnarray}
where $z$ is the spatial direction orthogonal to the $x-y$ plane and $\epsilon$ is the permittivity of the material surrounding the strange-metal layer.
The addition of this boundary term does not change the linearized equations of motion,
but it does change the boundary conditions for the field fluctuations \cite{zaanen2015,gran2018}.
 
We restrict the current to the $x-y$ plane where the strange-metal layer is assumed to be located, which gives us the following equation for the current
\begin{equation}
    J^{\mu}(\textbf{x},z,t) =  J^{\mu} (\textbf{x},t)\delta(z), 
\end{equation}
but let the Coulomb interactions, i.e., the photons, live in three dimensions. Then we obtain, after Fourier transforming and integrating out the photon field \cite{mauri2019}, the following effective boundary action
\begin{eqnarray}\label{eq:1layeraction}
   S = \frac{1}{2} \int \frac{d\omega d^2q}{(2\pi)^3} J^{\mu}(-\omega,-{\bf q})
   \chi_{\mu\nu}^{-1} J^\nu(\omega,{\bf q}),
\end{eqnarray}
where $\chi_{\mu\nu}$ is the current-current response function and is given by
\begin{equation}
    \chi_{\mu\nu}^{-1} \equiv \Pi_{\mu\nu}^{-1} +\frac{e^2\eta_{\mu\nu}}{2\epsilon q}.
\end{equation}
Since we are in a condensed-matter system, we are only interested in the density-density response, which is the $00$-component. In principle, and if desired, the density-current response $\chi_{0i}$ and the current-current response $\chi_{ij}$ can also be obtained. Here we take only the $00$-component and then obtain the density-density response function
\begin{equation}
    \chi(\omega,{q}) = \frac{\Pi(\omega,{q})}{1-\frac{e^2\Pi(\omega,{q})}{2\epsilon {q}}}.
\end{equation}
We thus see that we obtain the effects of the Coulomb potential $e^2/2\epsilon {q}$ in a similar manner as seen in the Random Phase Approximation (RPA) except that $\Pi(\omega,{q})$ is not a non-interacting response function, but contains interaction effects via the use of the Gubser-Rocha model. The Coulomb potential is taken independent of frequency, which means we neglected retardation effects due to the assumption that $v_F \ll c$ for our non-relativistic system. 
\begin{figure}[h!]
        \centering
        \includegraphics[width=0.5\textwidth]{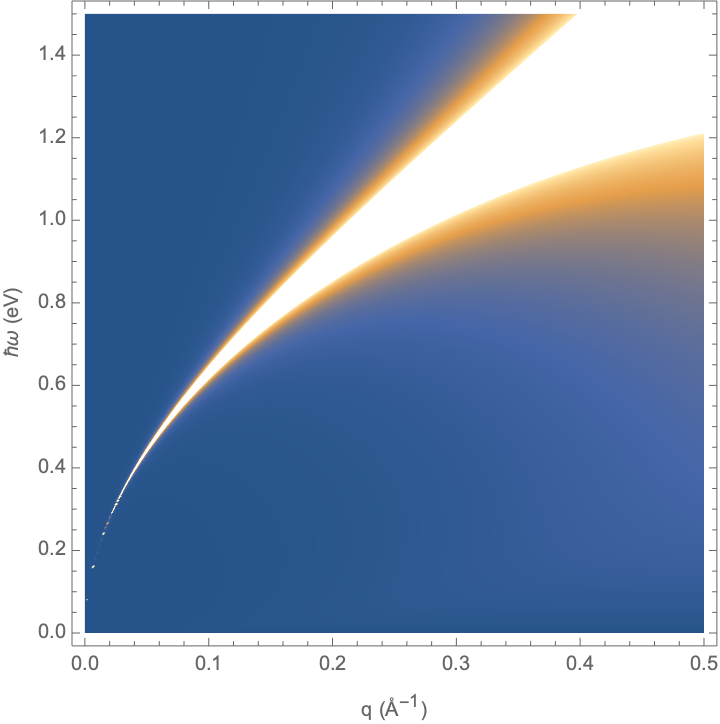}
        \caption{Density-density spectral function of a single layer with strong short-range interactions and long-range Coulomb interactions. We see a square-root plasmon mode in the density-density response, with $\omega \propto \sqrt{q}$. The temperature is fixed at room temperature $T=293$ K.}
        \label{fig:spectral1layercoulomb}
\end{figure}

We have plotted the resulting density-density spectral function $-\chi''$ in Fig \ref{fig:spectral1layercoulomb}. We now clearly see a dispersion $\omega \propto \sqrt{q}$, as expected in a two-dimensional system. Using Eq. (\ref{strongt=0}) we obtain the following dispersion relation
\begin{equation}
    \omega = \sqrt{\frac{e^2\mathcal{D}}{2\epsilon}q + v_s^2 q^2 }.
\end{equation} 
For low momenta $q$ we indeed have a square-root plasmon mode and at large $q$ we recover the sound dispersion from the previous section.
Furthermore, we see that the width of the plasmon peak in the spectral density quickly approaches zero as the momentum approaches zero, which indicates that at long wavelengths the plasmons have a long lifetime.
We also still have a diffusive mode, but it is barely visible in this figure. This is due to the fact that the intensity of the plasmon mode is much greater than the diffusive mode. But the latter mode is essentially still the same as in the neutral response function.

\subsection{\label{sec:level2} Bilayer plasmons}
Next, we introduce another identical layer at a distance $a$. In this section we derive the density-density spectral function of this system of two layers. The dominant interaction between the layers is the Coulomb force \cite{hepting2018}. The double-trace deformation is now carried out with a different expression for the current. The density current $ J^\mu $ in real space is now
\begin{equation} 
    J^{\mu}(\textbf{x},z,t) = J_1^{\mu}(\textbf{x},t)\delta(z+a/2)+  J_2^{\mu}(\textbf{x},t)\delta(z-a/2).
\end{equation}
This describes two strange-metal layers parallel to the $x-y$ plane separated by $a$ along the $z$-axis and the current is restricted to the two layers. The strong short-range interactions have no effect on the other layer, so we only need to account for the long-range Coulomb interaction between the two layers. Integrating out again the photon field, we obtain the following expression for the inverse of the bilayer response
\begin{equation}\label{eq:bilayer1}
    \chi^{-1}_{\mu\nu} = 
    \begin{pmatrix}
        \Pi^{-1}_{\mu\nu } + \frac{e^2 \eta_{\mu \nu}}{2\epsilon q} && \frac{e^2 \eta_{\mu \nu}e^{-qa}}{2\epsilon q}\\
        \frac{e^2 \eta_{\mu \nu}e^{-qa}}{2\epsilon q} && \Pi^{-1}_{\mu\nu } + \frac{e^2 \eta_{\mu \nu}}{2\epsilon q} 
    \end{pmatrix} .
\end{equation}
As in the previous section, we take the $00$-component and then invert the $2\times 2$ matrix with layer indices to obtain the density-density response function of one of the layers, i.e., $\chi_{II} \equiv \chi_{00,II}$. Here $I,J$ are the layer indices and refer to the components of the matrix in Eq. (\ref{eq:bilayer1}). Note that the density-density response function of the bilayer as a whole equals $\chi \equiv \sum_{IJ} \chi_{IJ}$. 
\begin{figure}[h]
        \centering
        \includegraphics[width=0.48\textwidth]{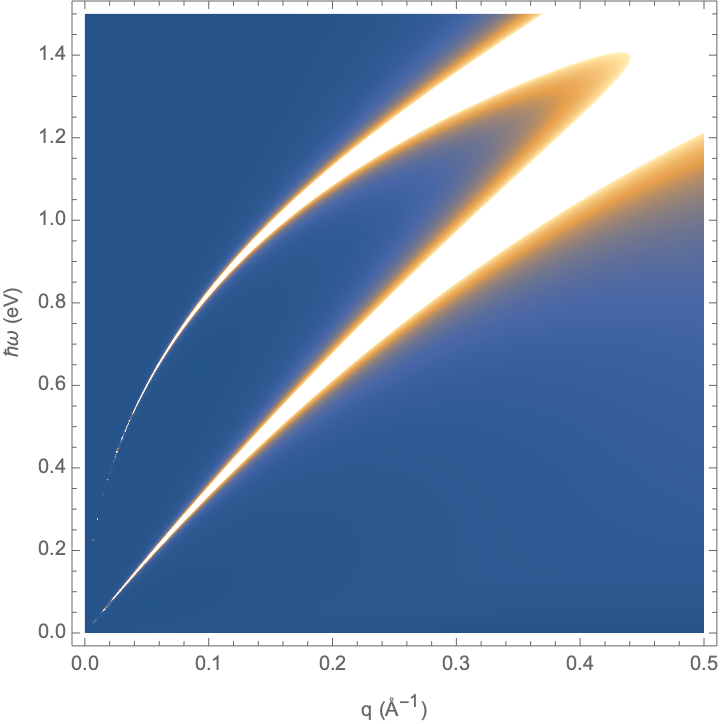}
        \caption{Diagonal part of the density-density spectral-function matrix of two layers with strong short-range interactions and long-range Coulomb interactions. The two layers are separated by $3.2$ \AA. This distance is due to the experimental setup that we model. We see the in-phase mode, $\omega \propto \sqrt{q}$ and the out-of-phase mode, $\omega \propto q$. The temperature is fixed at room temperature $T=293$ K.}
        \label{fig:spectral2layers}
\end{figure}

Then we can use this to plot the diagonal part of density-density spectral-function matrix $-\chi_{II}''$ in Fig. \ref{fig:spectral2layers}. Here we can see two modes. 
One mode has, as in the single-layer case, a square-root dispersion $\omega \propto \sqrt{q}$. This is the in-phase mode, whose behavior is similar as in the single-layer case. The width of this mode decreases very quickly as $q$ approaches zero. Then there is also another mode visible. This is a linear sound mode and this additional mode is the main difference between the bilayer and the single-layer case. It is called the out-of-phase mode, because the density fluctuations in adjacent planes are out of phase. Since there are no total charge fluctuations in this mode, we recover sound. We can again substitute the zero-temperature Gubser-Rocha response to obtain the dispersion relations
\begin{equation} \label{eq2layersdisp}
    \omega = \sqrt{\frac{e^2\mathcal{D}(1 \pm e^{-qa})}{2\epsilon}q + v_s^2 q^2  }.
\end{equation}
The plus sign is for the in-phase mode and the minus sign is for the out-of-phase mode. Notice that in the limit $a \rightarrow \infty$, both dispersion reduces to the single-layer case, which is as expected for two uncoupled layers. 

Next, we expand the dispersion for small $q$. The dispersion of the in-phase mode is
\begin{equation}
    \omega = \sqrt{\frac{e^2 \mathcal{D}q}{\epsilon}} + \mathcal{O}(q^{3/2}).
\end{equation}
To lowest order this is similar to the single-layer dispersion. The only difference is an additional factor of two under the square root. That is because the total density is twice as big as the single-layer density. In the dispersion this effectively doubles the Drude weight.
For the out-of-phase mode, using the minus sign in Eq. (\ref{eq2layersdisp}), we obtain the following expansion of the dispersion
\begin{equation}
    \omega = \sqrt{v_s^2+\frac{e^2\mathcal{D}a}{2\epsilon}} q + \mathcal{O}(q^2) .
\end{equation}
So the renormalized speed of sound of this mode is $\sqrt{v_s^2 +{e^2\mathcal{D}a}/{2\epsilon}}$. 
The width of this mode decreases less quickly and is wider over the whole range of $q$, compared to the in-phase mode. Again there is also a diffusive mode, which is not visible in this plot range because of the intensity of the plasmon modes. But it is the same as in the previous plots and has $\omega = -i D_dq^2$.

\section{\label{sec:level1}Layered strange metal}

\subsection{\label{sec:level2}Two layers per unit cell}

The next step is to stack such bilayers in an infinite periodic crystal. In this section we will derive the spectral function of this crystal of bilayers.
We take again the two layers in the unit cell to be separated by the distance $a$. Then we define the distance between the centers of the neighboring unit cells to be $l$. If $l=2a$ this case reduces to a periodic crystal of single layers all separated by $a$. We will discuss that special limit at the end of this section for completeness.

First, we derive the appropriate Coulomb potential matrix for the case of interest here. Thus we construct the current operator of the crystal
\begin{eqnarray} 
    J^{\mu}(\textbf{x},z,t) &=& \sum_{n \in \mathbb{Z}} J_1^{\mu}(\textbf{x},z,t)\delta(z-nl+a/2) \nonumber\\
    && + J_2^{\mu}(\textbf{x},z,t)\delta(z-nl-a/2).
\end{eqnarray}
We then Fourier transform this expression and integrate out the photon field. Then we obtain the Coulomb contribution to the effective boundary action for the currents as
\begin{widetext}
    \begin{equation} \label{potentialstack1}
        \Delta S_C = \frac{1}{2} \int \frac{d\omega d^2q}{(2\pi)^3}\int \frac{dq_z}{2\pi} \sum_{n,m} 
    \begin{pmatrix}
        J^\mu_1(-\omega,-{\bf q},nl-a/2)\\
        J^\mu_2(-\omega,-{\bf q},nl+a/2)
    \end{pmatrix}
    \cdot
    \frac{e^2}{\epsilon}\frac{e^{-iq_z(n-m)l}}{q^2+q_z^2}
    \begin{pmatrix}
        \eta_{\mu \nu}& \eta_{\mu \nu}e^{-qa} \\
        \eta_{\mu \nu}e^{-qa} & \eta_{\mu \nu}\\
    \end{pmatrix} 
    \cdot
    \begin{pmatrix}
          J^\nu_1(\omega,{\bf q},ml-a/2)\\
          J^\nu_2(\omega,{\bf q},ml+a/2)
    \end{pmatrix}.
    \end{equation}
\end{widetext}
This equation is the bilayer-crystal equivalent of Eq. (\ref{eq:1layeraction}). 
We can perform the integration over $q_z$ and Fourier transform the periodicity over $n$ and $m$ to the Bloch momentum $p$, i.e., 
$ J^\mu_1(\omega,{\bf q},nl-a/2)= ({l}/{2\pi}) \int_{-\pi/l}^{\pi/l} dp  J^\mu_1(\omega,{\bf q},p)e^{ip(nl-a/2)}$. 
The Bloch momentum is in the direction perpendicular to the layers, since the periodicity is in $n$ and $m$.
After substituting the discrete Fourier transform into Eq. (\ref{potentialstack1}) we obtain the desired result 
\begin{eqnarray} \label{potentialstack2}
    \Delta S_C &=& \frac{1}{2} \int \frac{d\omega d^2q}{(2\pi)^3}\int_{-\pi/l}^{\pi/l} \frac{ldp}{2\pi}
    \begin{pmatrix}
        J^\mu_1(-\omega,-{\bf q},-p)\\
        J^\mu_2(-\omega,-{\bf q},-p)  
    \end{pmatrix}
    \nonumber\\
    && \cdot \frac{e^2\eta_{\mu\nu}}{\epsilon} V(q,p) \cdot
    \begin{pmatrix}
          J^\nu_1(\omega,{\bf q},p) \\
          J^\nu_2(\omega,{\bf q},p)
    \end{pmatrix},
\end{eqnarray}
with the following expression for the $2 \times 2$ matrix $V$, with the same form as for a bilayered electron-gas \cite{griffin1989,schulte1999},
\begin{widetext}
    \begin{equation}
        V(q,p) = \frac{1}{2q(\cosh{ql}-\cos{pl})}
    \begin{pmatrix}
         \sinh{ql}&(\sinh{q(l-a)}+e^{-ipl}\sinh{qa})e^{-ipa}\\
        (\sinh{q(l-a)}+e^{ipl}\sinh{qa})e^{ipa}&\sinh{ql}
    \end{pmatrix}.
    \end{equation} 
\end{widetext}

As in the previous section we thus construct the inverse of the response function as 
\begin{equation} \label{response2}
    \chi^{-1}_{\mu\nu } = \Pi^{-1}_{\mu\nu } + \frac{e^2 \eta_{\mu\nu}}{\epsilon}V.
\end{equation}
We are interested in the density-density response, so we take the 00-component, with the same reasoning as in the previous sections.  
\begin{figure*}
    \centering
    \includegraphics[width=0.9\textwidth]{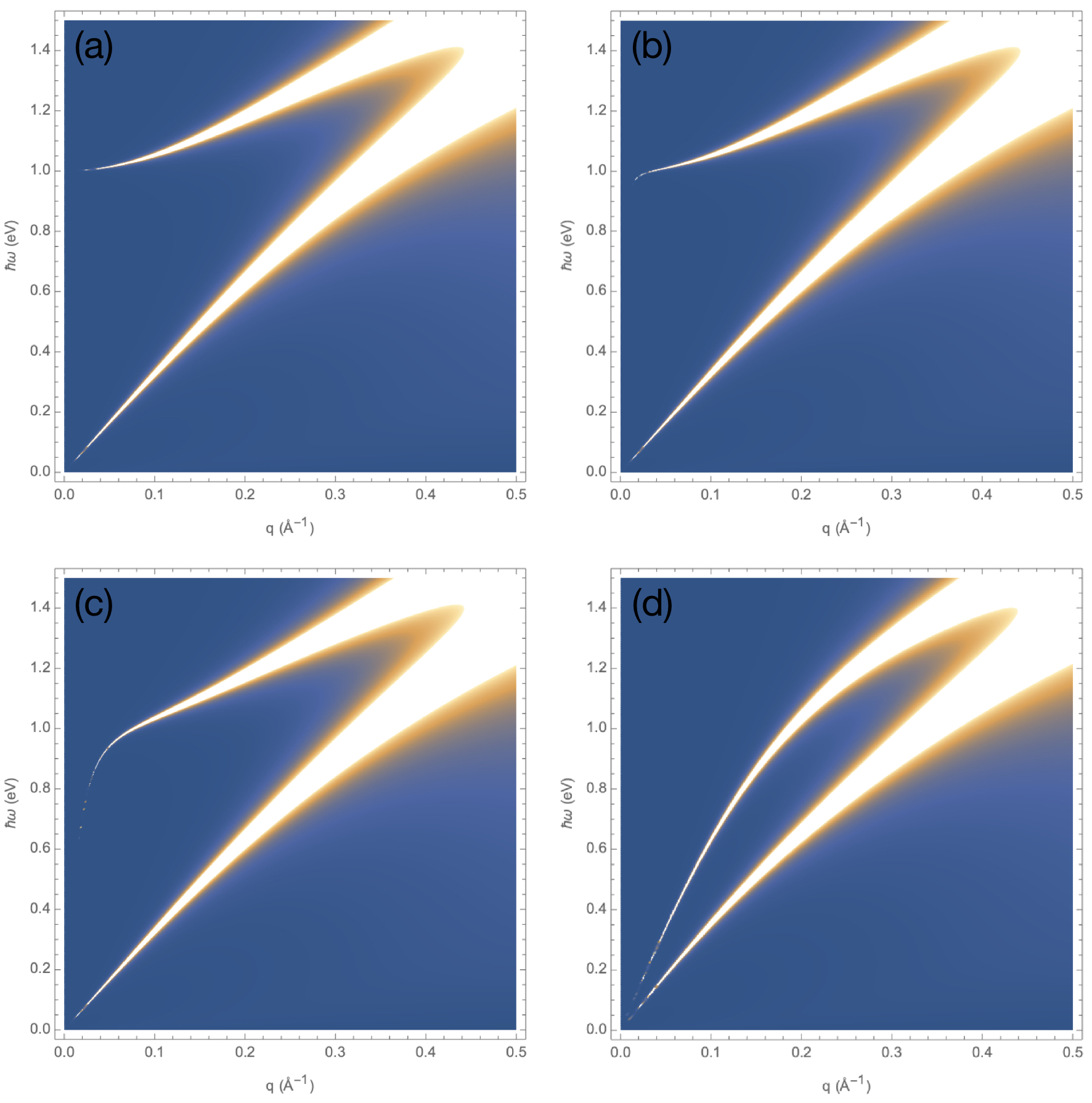}
    \caption{Diagonal part of the density-density spectral-function matrix of a crystal of bilayers. Each pair of layers is separated by $a=3.2$ \AA \, and the size of each unit cell is $l=15.4$ \AA. Each subfigure has a different value of $pl$. In (a), $pl=0$, the in-phase plasmon mode is gapped as expected for a three-dimensional system and we observe in addition an out-of-phase sound mode because we have two layers per unit cell. In (b)-(d), $pl$ is $\pi/50$, $\pi/10$ and $\pi$, respectively. Here the out-of-phase sound mode is almost unaffected by the out-of-plane momentum $p$, whereas the in-phase plasmon mode is no longer gapped and has obtained also an acoustic behavior at long wavelengths. The temperature is fixed at room temperature $T = 293$ K.}
    \label{fig:spectral2layersstack}
\end{figure*}

We have plotted the diagonal part of the density-density spectral-function matrix $-\chi_{II}''$ in Fig. \ref{fig:spectral2layersstack} for multiple values for the Bloch momentum $p$. In each plot there are two modes visible as we are dealing with periodicity of bilayers in the out-of-plane direction and we thus obtain a periodic band structure for the in-phase and the out-of-phase modes. In Fig. \ref{fig:spectral2layersstack}(a) we have shown the case $p=0$.  In this case, the in-phase mode is gapped with an energy of $1.0$ eV. This is the plasma frequency $\omega_{pl}$ for Bi-2212. For $p \neq 0$ the in-phase mode is not gapped, however. We see in Fig. \ref{fig:spectral2layersstack}(b) that the in-phase mode approaches $1.0$ eV for smaller momenta but ultimately bends down to zero at the longest wavelengths and obtains an acoustic character with a speed of sound that strongly depends on the Bloch momentum. Indeed, in the other subplots, for higher values of $p$, the mode becomes less steep for low momenta. For $p=\pi/l$ the speed of sound of this mode is the lowest and close to the speed of sound of the Gubser-Rocha model for the single layer. 
In contrast, we observe that the out-of-phase mode has no significant $p$-dependence. This makes sense physically, because there is hardly any Coulomb coupling between the different bilayers in this case, since they are charge neutral.
\begin{figure*}
     \includegraphics[width=0.9\textwidth]{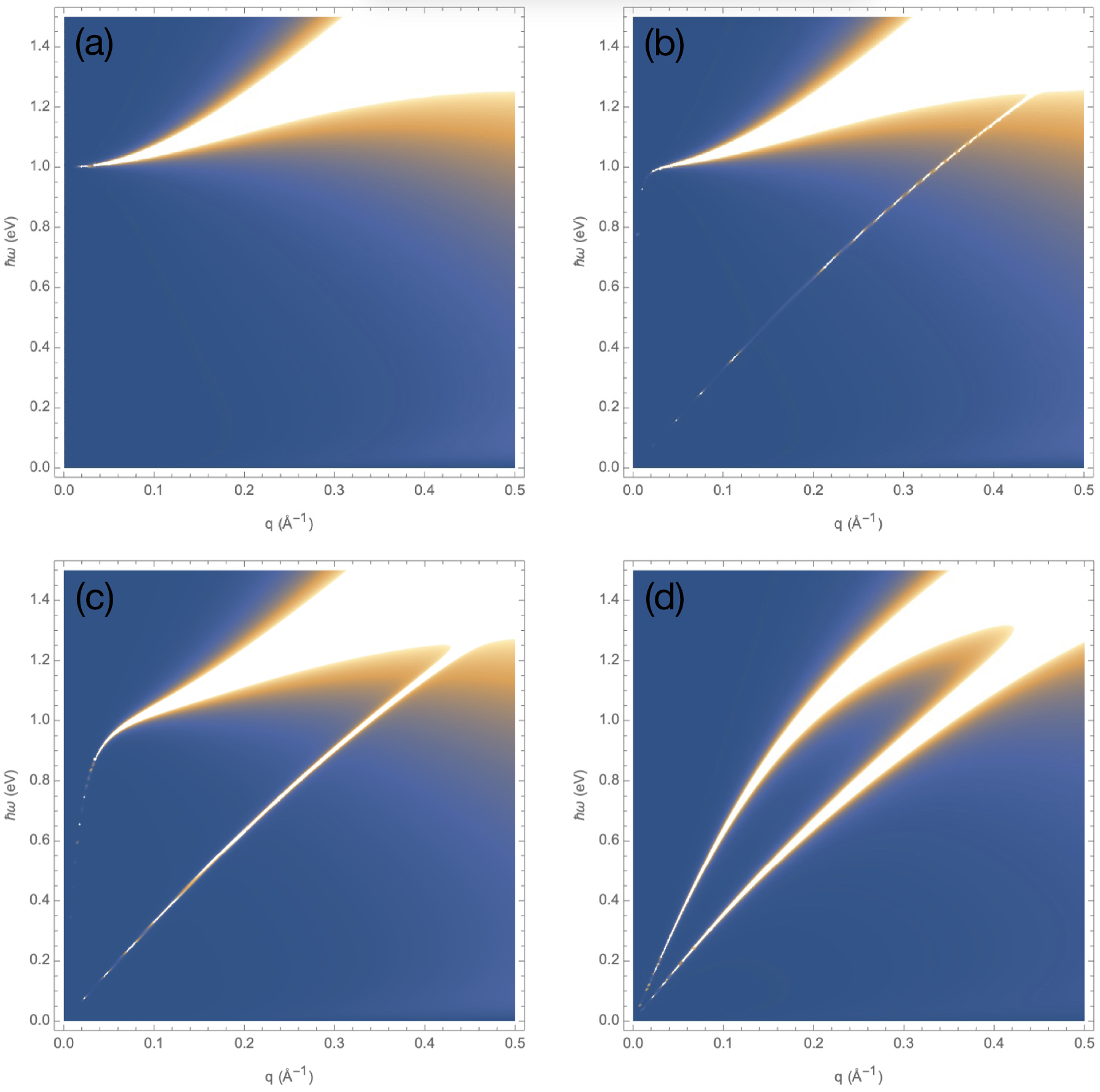}
      \caption{
      Total density-density spectral function of a crystal of bilayers. Each pair of layers is separated by $a=3.2$ \AA, and the size of each unit cell is $l=15.4$ \AA. In (a), $pl=0$, the in-phase plasmon mode is gapped as expected for a three-dimensional system. Compared with Fig. \ref{fig:spectral2layersstack} the out-of-phase mode is cancelled completely. In (b)-(c), $pl$ is $\pi/50$ and $\pi/10$, respectively. Here the out-of-phase sound mode appears, with intensity and width increasing as $p$ increases. The in-phase plasmon mode is no longer gapped and has obtained also an acoustic behavior at long wavelengths. In (d), $pl=\pi$, the plot shows the lowest speed of sound of the in-phase mode. The temperature is fixed at room temperature $T = 293$ K.}
    \label{fig:logspectral2layers}
\end{figure*}

Then in Fig. \ref{fig:logspectral2layers} we have plotted the total density-density spectral function. It is defined as the sum of all components of the matrix defined in Eq. (\ref{response2}). While the in-phase mode intensity peaks at $p=0$, the out-of-phase mode vanishes for $pl=0$ and its intensity increases for larger $p$ and peaks at $p=\pi/l$.
We use the total density-density spectral function in later sections to define the total conductivity and the loss function. 

As in previous sections we substitute the zero-temperature neutral response to analytically compute the dispersion. We obtain the following expression
\begin{equation}
    \omega = \sqrt{ \frac{e^2\mathcal{D}q}{2\epsilon}\frac{\sinh{ql}\pm |\sinh{q(l-a)}+ e^{ipl}\sinh{qa}|}{\cosh{ql}-\cos{pl}} + v_s^2q^2}.
\end{equation}
Again, the minus sign corresponds to the out-of-phase mode and the plus sign to the in-phase mode. The first thing to check is if this dispersion reduces to the bilayer case in the limit $l\rightarrow \infty$. This is indeed true, because in this limit both $\sinh{ql}$ and $\cosh{ql}$ become equal to $e^{ql}/2$ and the complicated fraction under the square root indeed exactly reproduces the result of the bilayer case in Eq.  (\ref{eq2layersdisp}). We also know that for $p=0$ there is a gapped mode. Using the above equation we can derive an equation for the associated plasma frequency. Taking the limit $q\rightarrow0$ for the in-phase mode and $p=0$, we obtain 
\begin{equation}
    \omega_{pl} = \sqrt{\frac{2e^2\mathcal{D}}{l\epsilon}}.
\end{equation}
Here the plasma frequency is defined in terms of the two-dimensional Drude weight $\mathcal{D}$. But we can also write it in terms of the three-dimensional Drude weight 
$\mathcal{D}_{3D} = {2\mathcal{D}}/{l}$, which shows 
that the plasma frequency equals the familiar result $\omega_{pl}= \sqrt{{e^2\mathcal{D}_{3D}}/{\epsilon}}$.
Using these equations we rewrite the dispersion 
\begin{equation}
    \omega = \sqrt{\omega_{pl}^2\frac{ql}{4}\frac{\sinh{ql}\pm |\sinh{q(l-a)}+ e^{ipl}\sinh{qa}|}{\cosh{ql}-\cos{pl}} + v_s^2q^2 },
\end{equation}
from which we can derive an expression for the two speeds of sound  for $p \neq 0$. Namely, we find
\begin{equation}
    {v}_\pm(p) = \sqrt{v_s^2 + \frac{\omega_{pl}^2 l}{4} \frac{l \pm |l+(e^{ipl}-1)a|}{1-\cos{pl}} }.\\
\end{equation}

\subsection{\label{sec:level3}Derivation of parameters}
In this section we give the values of the various constants in Eq. (\ref{strongresponse}), the Gubser-Rocha response function, and explain how the in principle unknown prefactor $N_G$ of the gravitational action can be determined using the plasma frequency. We start with the thermodynamic equation of state for the electron density inside each layer obtained from the holographic dictionary as \cite{mauri2022}
\begin{equation}
    n = \frac{N_G'}{\sqrt{3}}\left( \frac{\mu}{\hbar v_F}\right)^2\sqrt{1+\frac{1}{3}\left(\frac{k_B T}{\mu} \right)^2},
\end{equation}
where $N_G'=N_G/\Tilde{e}$ and $\Tilde{e}$ is the dimensionless charge. We rewrite this such that we have a formula for the chemical potential $\mu$ in terms of the temperature $T$ and the electron density $n$ as 
\begin{equation}
    \mu = \sqrt{\frac{\sqrt{(k_B T)^4+108 \left(\frac{\hbar v_F \sqrt{n}}{\sqrt{N_G'}}\right)^4}-(k_B T)^2}{6}} .
\end{equation}
Then we expand this near zero temperature to obtain
\begin{equation}
    \mu = \frac{3^{1/4}\hbar v_F \sqrt{n}}{\sqrt{N_G'}} -\frac{
    \sqrt{N_G'}}{4 \times 3^{5/4}} \frac{(k_B T)^2}{\hbar v_F \sqrt{n}} +\mathcal{O}(T^4) .
\end{equation}

Following the derivation of Mauri and Stoof \cite{mauri2022}, we use the result for the Drude weight of the Gubser-Rocha theory as $\mathcal{D}= {N_G'} {\mu(n,T)}/{\sqrt{3}\hbar^2}$.  
Now, we also use the relation between the Drude weight and the plasma frequency, which we can then use to relate $N_G'$ to the plasma frequency. We obtained previously for the Drude weight
\begin{equation} \label{drude1}
    \mathcal{D} = \omega_{pl}^2 \frac{l \epsilon}{2 e^2} .
\end{equation}
The plasma frequency is essentially temperature independent, so we know that the Drude weight should also be temperature independent, since the other quantities in Eq. (\ref{drude1}) are as well. 
Making use of this observation we then derive the expansion for $N_G'$ up to second order in temperature with the result that
\begin{eqnarray}
    N_G' &=& \left( \frac{3^{1/4}\hbar}{v_F \sqrt{n}} \frac{\omega_{pl}^2 l\epsilon}{2e^2}\right)^2 +\frac{1}{2\times 3^{3/2}} \left( \frac{3^{1/4}\hbar }{v_F \sqrt{n}}  \frac{\omega_{pl}^2 l\epsilon}{2e^2}\right)^4 \nonumber\\ 
    && \times  \left( \frac{k_B T}{\hbar v_F \sqrt{n}}\right)^2 
    + \mathcal{O}(T^4).
\end{eqnarray}
We have already computed the Drude weight, so now we can also derive the hydrodynamic compressibility $\chi$ using a relation obtained previously \cite{mauri2022}. Up to lowest order in temperature $\chi= \mathcal{D}/v_s^2$. While at quadratic order in temperature the difference is given by
\begin{equation}
    \chi - \frac{\mathcal{D}}{v_s^2} = 5.12 (N_G')^{3/2} \left( \frac{k_B T}{\hbar v_F \sqrt{n}}\right)^2 \frac{\sqrt{n}}{\hbar v_F}.
\end{equation}
The factor $({N_G'})^{3/2}$ is due to the fact that the extensive parameters $\chi$ and $\mathcal{D}$ are multiplied by $N_G'$ and the density should be divided by $N_G'$ as it is obtained from the square of the chemical potential. Since we can compute the Drude weight, we can now explicitly compute the compressibility too. Then we still have only the two diffusion coefficients left, characterizing the charge diffusion and the sound diffusion. These parameters are inversely proportional to the density $n$, which means they should be multiplied by $N_G'$ and are thus equal to 
\begin{align}
    D_s = \frac{1}{6\sqrt{3}} {N_G'} \frac{k_B T}{\hbar n},\\
    D_d = \frac{4\pi}{\sqrt{3}} {N_G'} \frac{k_B T}{\hbar n}.
\end{align}

For Bi-2212 that is of special interest to us here, we have used the following material parameters: $\hbar \omega_{pl}=1.0$ eV, $l=15.4$ \AA, $a=3.2$ \AA, $\epsilon/e^2 = 4.5 \times 55.263 \times 10^{-4} $eV$^{-1} $\AA$^{-1}$, $v_F=2.28\times 10^{5}$ m\! s$^{-1}$, $v_s=1.73\times 10^{5}$ m\! s$^{-1}$, and $n=6.25 \times 10^{18} $ m$^{-2}$. We used these values to plot all the density-density spectral functions. Note that in particular we have $N_G' \simeq 0.45$ at zero temperature.

\begin{figure*}
    \centering
    \includegraphics[width=\textwidth]{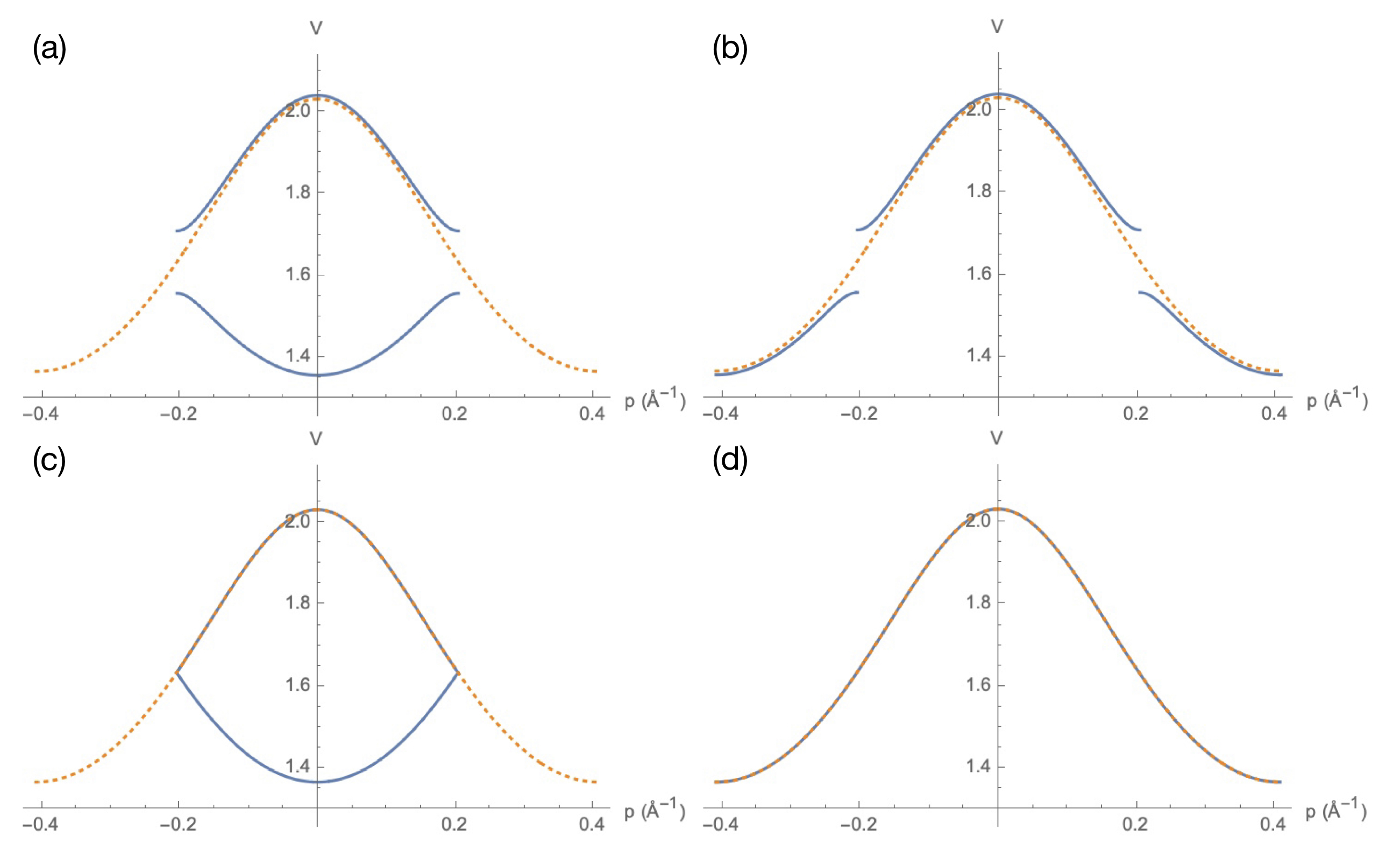}
   \caption{The quantity $V_{\pm}$ in blue is compared with $V_{\text{singlelayer}}$ in dotted orange for a fixed value of $q=0.3$ \AA$^{-1}$ and as a function of the Bloch momentum $p$. The two upper figures are plotted with value $a=0.45l$, the two bottom figures for $a=l/2$.
    The minus sign of $V_{\pm}$ corresponds to the out-of-phase mode with a lower value and the plus sign to the in-phase mode with the higher value. In (a) and (c), we have plotted the bilayer potential for $p \in [-\pi/l,\pi/l]$ and then in (b) and (d) we have periodically extended the minus sign solution to $[-2\pi/l,-\pi/l] \cup [\pi/l,2\pi/l]$. In this plot $l=15.4$ \AA.  } 
    \label{fig:limit}
\end{figure*}
\begin{figure*}
    \centering
    \includegraphics[width=0.9\textwidth]{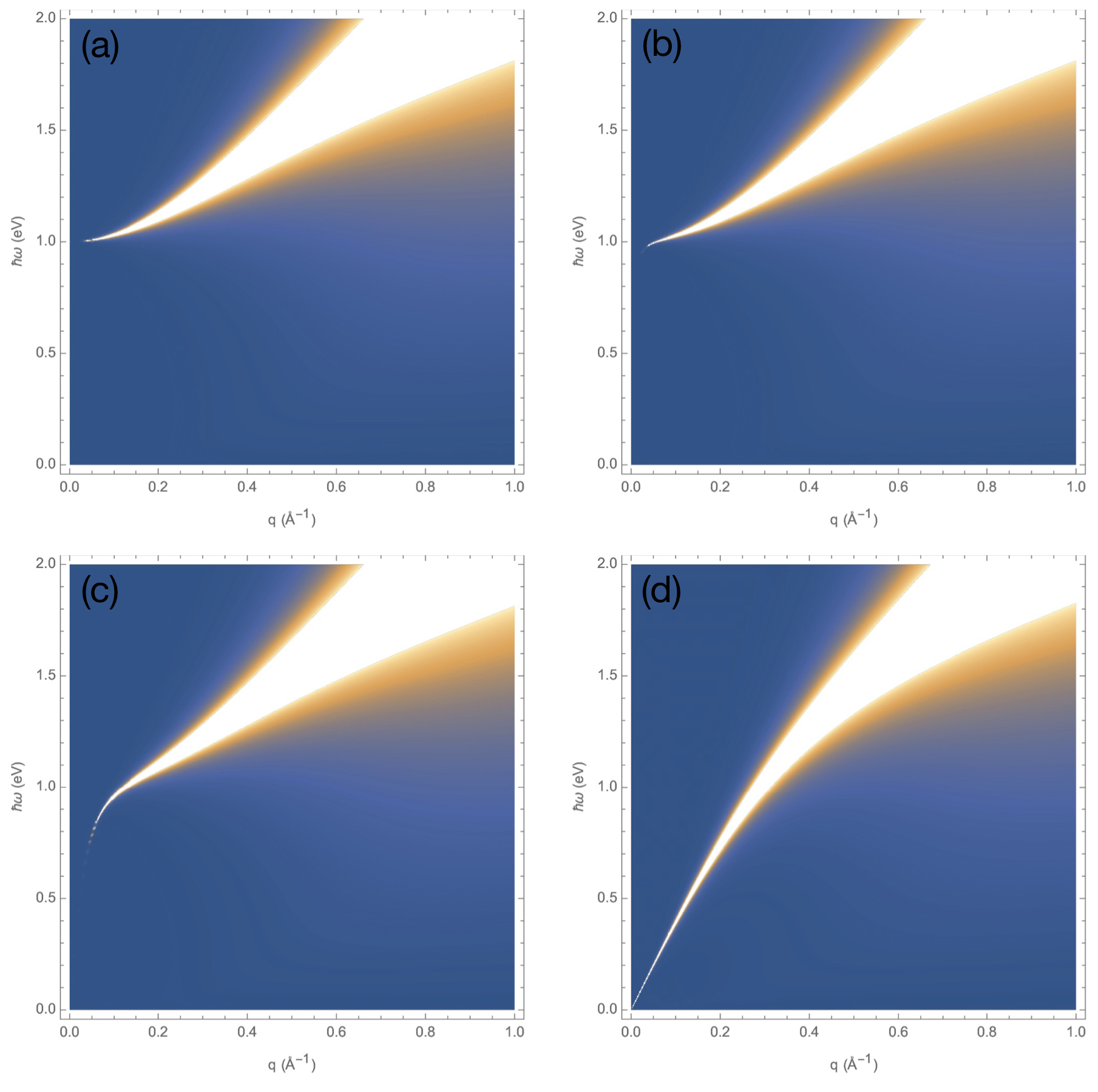}
     \caption{Density-density spectral function of a crystal of single layers. The distance between layers $l/2 =7.7$  \AA. In (a), $pl=0$, there is a gapped plasmon mode. In (b)-(d), $pl$ is $\pi/50$, $\pi/10$ and $\pi$, respectively. Here is an acoustic plasmon mode visible at long wavelengths, as in the crystal of bilayers. The temperature is fixed at room temperature $T=293$ K.}
    \label{fig:spectral1layer}
\end{figure*}

\subsection{\label{sec:level2} One layer per unit cell}
In the limit $a \rightarrow l/2$ the bilayer model reduces to a layered crystal with all neighboring layers having an equal distance $l/2$ between them. There is also a bismuth-based cuprate which has this structure, given by Bi$_2$Sr$_2$CuO$_{6+x}$, also known as Bi-2201. The response function of this crystal has been computed before \cite{mauri2019}. In that case the intralayer physics was different, and did not represent a strange metal, but the long-range Coulomb force is treated in the same way. We show now that our response function reduces to this case of a single layer per unit cell in the limit $a \rightarrow l/2$. This is to corroborate the expression for the crystal of bilayers. From Eq. (\ref{response2}) we can deduce that the only relevant quantity related to the Coulomb potential matrix in the dispersion is $V_\pm \equiv V_{11}\pm \sqrt{V_{12}V_{21}}$. This can be compared to the same quantity of a crystal of single layers given by \cite{mauri2019}
\begin{equation}
    V_{\text{singlelayer}}= \frac{\sinh{\frac{ql}{2}}}{2q(\cosh{\frac{ql}{2}}-\cos{\frac{pl}{2}})} ,
\end{equation}
with the distance between the layers taken equal to $l/2$. Our expression for the bilayer case is
\begin{equation}
    V_{\pm} = \frac{\sinh{ql}\pm |\sinh{q(l-a)} + e^{ipl}\sinh{qa}|}{2q(\cosh{ql}-\cos{pl})} .
\end{equation}
Here $a$ is the distance between the layers in the unit cell and $l$ is the distance between the unit cells. We plot in Fig. \ref{fig:limit} $V_{\pm}$ for a fixed value of $l$ and change the value of $a$. Then we compare this with the single-layer result.

In Fig. \ref{fig:limit}(a) and \ref{fig:limit}(b) we have chosen $a=0.45 l$, which means that the two layers in the unit cell are relatively far apart and the system approaches the limit of a crystal of equidistant layers. In Fig. \ref{fig:limit}(a) we have plotted both quantities $V_\pm$ for $p \in [-\pi/l,\pi/l]$. Then in Fig. \ref{fig:limit}(b)  we have extended $V_-$ to $[-2\pi/l,-\pi/l] \cup [\pi/l,2\pi/l]$. This is to show that $V_{\pm}$ is already almost equal to the single-layer result in the extended-zone scheme. Then in Fig. \ref{fig:limit}(c)  we have plotted the limiting case of $a=l/2$. We see that at $p=\pm\pi/l$ the two solutions are exactly matched to each other. Finally, in Fig. \ref{fig:limit}(d) we extend the minus sign solution and see that $V_{\pm}$ is equal to the single-layer potential for $l/2$, as expected. This means that in the limit $a\rightarrow l/2$ the crystal of bilayers reduces to the crystal of single layers. Which reinforces that the expression for the bilayer Coulomb potential matrix is correct. \\

Finally, we also plot the density-density spectral function of the crystal of single layers in Fig. \ref{fig:spectral1layer}. Comparing these figures with previous work \cite{mauri2019}, we see that this spectral function also has a gapped mode for $p=0$. The plasma frequency is the same, $\hbar \omega_{pl}=1.0$ eV. This is because the three-dimensional density of electrons has not changed, since there still is one layer per $l/2$ in the z-direction. And that is what determines the plasma frequency. Mathematically the reason is that $2n/l = n/(l/2)$. In the other subplots the behavior is as expected, it is similar to the in-phase mode of Fig. \ref{fig:spectral2layersstack}. For ${pl}/{2} =\pi/50$, the mode approaches $1.0$ eV but then quickly goes to zero as $q \rightarrow 0$. For increasing $p$ the mode becomes less steep for small $q$ until for ${pl}/2=\pi$ the lowest speed of sound is reached.
In the limit $T\rightarrow 0$ we can again derive the dispersion relation
\begin{equation}
    \omega = \sqrt{\frac{e^2\mathcal{D}q}{2\epsilon}\frac{\sinh{(\frac{ql}{2})}}{\cosh{(\frac{ql}{2})}-\cos{(\frac{pl}{2})}}+ v_s^2q^2 }.
\end{equation}
The plasma frequency is exactly the same as in the two-layer case, since the three-dimensional density is kept constant. So the dispersion relation becomes
\begin{equation}
    \omega = \sqrt{\omega_{pl}^2 \frac{ql'}{2}\frac{\sinh{(ql')}}{\cosh{(ql')}-\cos{(pl')}} + v_s^2q^2 },
\end{equation}
where $l' = l/2$ is the distance between the layers. The plasma frequency is given by $\omega_{pl} = \sqrt{{e^2\mathcal{D}}/{l'\epsilon}}$. Using this equation we can derive the renormalized speed of sound
\begin{equation}
    {v(p)} = \sqrt{ v_s^2 + \omega_{pl}^2 \frac{l'^2/2}{1-\cos{(pl')}} } ,
\end{equation}
confirming that ${pl}/{2}= pl' = \pi$ gives the lowest speed of sound. The above equation is not valid exactly for $p=0$, of course. Besides the plasmon mode, this spectral function also contains a diffusive mode. Although this mode is not visible due its low intensity compared to the plasmon mode. 

Finally, we wish to emphasize that the spectral function in Fig. \ref{fig:spectral1layer} is in accordance with a number of resonant inelastic X-ray scattering (RIXS) studies on the strange-metal phase of cuprates with one CuO$_2$ layer per unit cell. In two of these RIXS studies on the electron-doped cuprate La$_{2-y}$Ce$_y$CuO$_{4+x}$ (LCCO) \cite{hepting2018} and the hole-doped cuprates La$_{2-y}$Sr$_y$CuO$_{4+ x}$ (LSCO) and Bi$_2$Sr$_{1.6}$La$_{0.4}$CuO$_{6+x}$ (Bi-2201) \cite{nag2020}, an acoustic plasmon dispersion is measured. The corresponding RIXS intensity maps are qualitatively similar to Fig. \ref{fig:spectral1layer}, with the corresponding nonzero values of $pl'$. The density-density spectral function in Fig. \ref{fig:spectral1layer} is calculated using an approach that is appropriate for strange metals, namely using the Gubser-Rocha model. This validates the conclusion that the acoustic branches which are measured in RIXS studies can be attributed to an acoustic plasmon.

\section{\label{sec:level1}Conductivity}
In this section we consider for completeness also the conductivity of the bilayer crystal. First, we obtain the following formula for the total density-density response function in the long-wavelength limit $q,p \rightarrow 0$
\begin{equation} \label{eq:cond1}
      \chi(\omega, {q}, p) = \frac{2\mathcal{D}q^2}{\omega^2-\omega_{pl}^2},
\end{equation}
where $\chi = \sum_{IJ} \chi_{IJ}$ is the total density-density response function of the bilayer crystal.
We can rewrite this by factoring $i\omega$ out in the denominator, leading to
\begin{equation}
  \chi(\omega, {q}, p) = 
    \frac{2\mathcal{D}q^2}{-i\omega}\frac{1}{i\omega +\frac{\omega_{pl}^2}{i\omega}}.
\end{equation}
Next, we observe that the second denominator has the form recognizable from the continuity equation for the electron density, together with both Ohm's law and Gauss's law, namely
\begin{equation}
    \left(- i\omega + \frac{ \sigma(\omega)}{\epsilon} \right) n =0 ,
\end{equation}
which means that the conductivity is 
\begin{equation} \label{eq:cond2}
    \sigma(\omega)= \frac{e^2 \mathcal{D}_{3D} }{-i\omega}. 
\end{equation}
Note that the same result can also be obtained directly from the `neutral' in-plane conductivity as 
\begin{equation}
 \sigma(\omega)= \frac{2}{l} e^2 \lim_{q \rightarrow 0} \frac{i\omega}{q^2} \Pi(\omega,q) = \frac{2e^2}{l}  \frac{\mathcal{D}}{-i\omega},
\end{equation}
which is as expected physically since Coulomb interactions do not affect the acceleration of the total momentum due to the applied electric field.

In first instance the real part of the above expression leads to a delta function centered around $\omega=0$ that signals the absence of momentum relaxation in our theory. But in an experiment there is typically disorder in the sample. We can incorporate this using the Planckian dissipation appropriate for the cuprates \cite{legros2019}, by performing the replacement $\omega \rightarrow \omega + \frac{i}{\tau}$ in the right-hand side. Planckian dissipation gives us the following expression for the relaxation time $\tau$
\begin{equation}
    \frac{\hbar}{\tau} = \alpha k_B T,
\end{equation}
with $\alpha$ a material parameter. For Bi-2212, it is approximately $1.1 \pm 0.3$ \cite{legros2019}. The fact that the dissipation rate is linear in temperature aligns with the fact that the diffusion constants in the strong short-range response function are also linear in temperature. So this is consistent with our use of the Gubser-Rocha model for the strange-metal phase. After introducing Planckian dissipation in the above manner, Eq. (\ref{eq:cond2}) obtains the Drude form with the dc-conductivity inversely proportional to temperature. Which means that the resistivity is linear in temperature, as required for strange metals.

\section{\label{sec:level1} Loss Function}
We are now in the position to discuss the so-called loss function that can be measured experimentally in transmission EELS measurements. We incorporate the non-zero in-plane momentum resolution in the experiments by defining the following average loss function
\begin{widetext}
    \begin{equation}
    L(\omega,q_0,p) = \frac{2}{\Delta q^2} \frac{1}{e^{-\left(\frac{q_0}{\Delta q}\right)^2}+\sqrt{\pi} \frac{q_0}{\Delta q} \left(1+ \text{erf}\frac{q_0}{\Delta q}\right)} \int_{0}^{\infty} dq q e^{-\left(\frac{q-q_0}{\Delta q}\right)^2} \text{Im} \left[-\frac{ \chi(\omega+\frac{i}{\tau}, {q},p)}{q^2} \right].
\end{equation}
\end{widetext}

In this formula $\chi$ is again the density-density response function of interest to us, and we take the imaginary part, which shows that transmission EELS measures essentially the density-density spectral function. We have incorporated disorder in the same fashion as in the previous section and performed the replacement $\omega \rightarrow \omega + \frac{i}{\tau}$. Then we average with a Gaussian distribution centered around $q_0$, denoting the measured in-plane momentum transfer, and with a width $\Delta q$ that represents the experimental momentum resolution. 
\begin{figure}
    \includegraphics[width=0.5\textwidth]{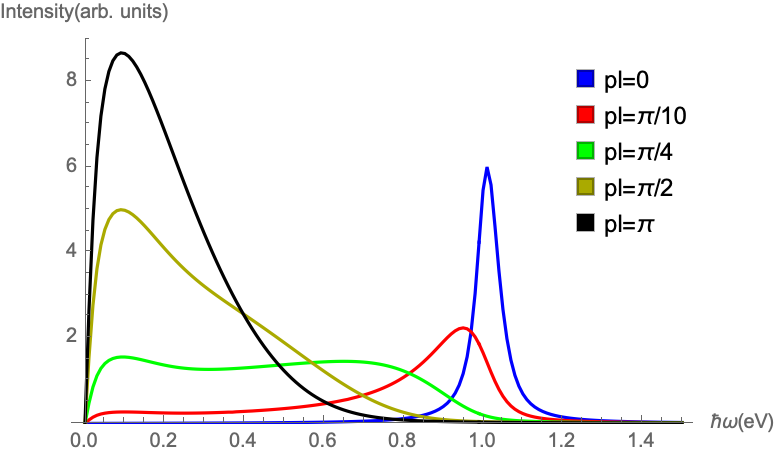}        
     \caption{The loss function for $q_0=0$. The uncertainty is characterized by $\Delta q=0.05$ \AA$^{-1}$. Multiple values of $p$ are used, indicated in the legends, and we also describe Planckian dissipation with $\alpha=1.1$. There is strong $p$ dependence in this case.} 
     \label{fig:lf1}
\end{figure}
\begin{figure}
    \centering
    \includegraphics[width=0.5\textwidth]{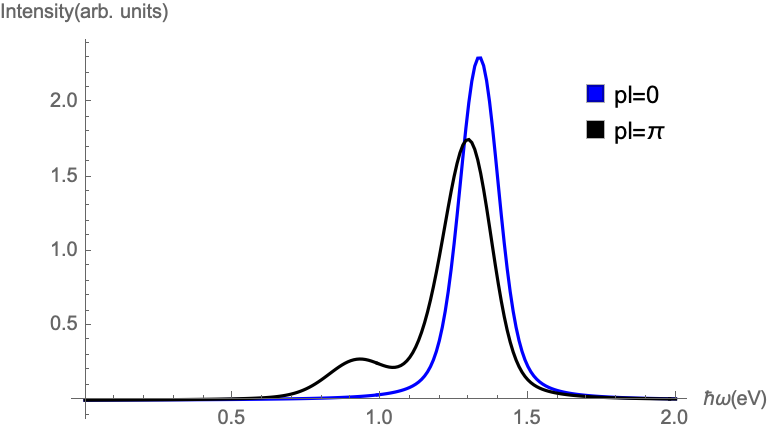}   
    \caption{The loss function for $q_0=0.3$ \AA$^{-1}$. The uncertainty is characterized by $\Delta q=0.05$ \AA$^{-1}$. Multiple values of $p$ are used, indicated in the legends and we also include Planckian dissipation with $\alpha=1.1$. Compared with Fig. \ref{fig:lf1} there is little dependence on $p$. We have only plotted the outermost values of $p$, to make the figure more clear. We also observe that the energy of the peak has increased compared to Fig. \ref{fig:lf1}. The smaller peak visible for $pl=\pi$ is due to the out-of-phase mode.}
    \label{fig:lf2}
\end{figure}

In Fig. \ref{fig:lf1} we have plotted the loss function $L(\omega,q_0,p)$ for $q_0=0$ and $\Delta q=0.05$ \AA$^{-1}$. This loss function is strongly dependent on $p$. For $p$ close to $0$ we see a peak in the intensity around $1.0$ eV. And as $p$ increases the peak widens and decreases in intensity. Then for $pl=\pi$ we see that there is a peak around $0.15$ eV, which corresponds approximately to $2\hbar v_s \Delta q$. In Fig. \ref{fig:lf2} the in-plane momentum is increased to $q_0=0.3$ \AA$^{-1}$ and the $p$ dependence has almost disappeared. The energy of the plasmon peak has increased to around $1.3$ eV. 

\begin{figure}[h!]
        \centering
        \includegraphics[width=0.5\textwidth]{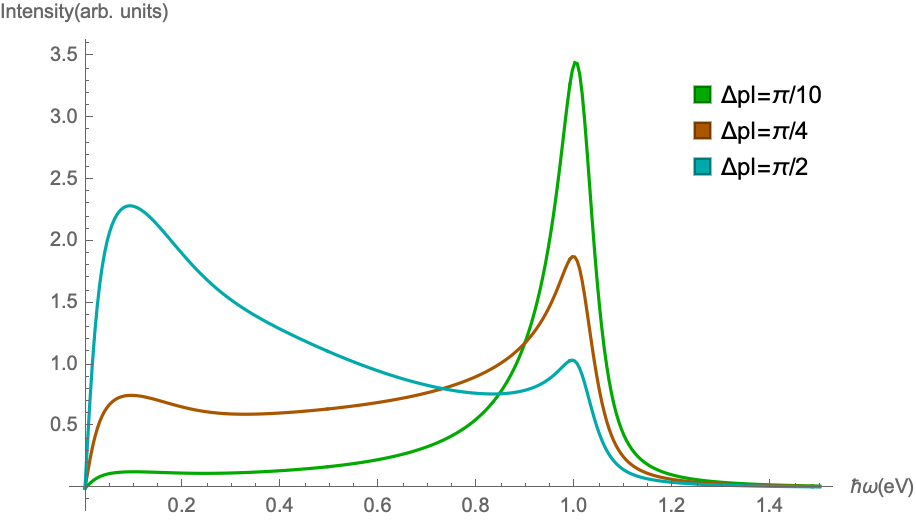}
        \caption{The dependence of the average loss function on the out-of-plane momentum resolution for $q_0=0$, $\Delta q=0.05$ \AA$^{-1}$, $p_0=0$, and $\alpha=1.1$. }
        \label{fig:lf3}
\end{figure}

In these plots we have assumed a single value of $p$, but an experiment could also have uncertainty in the out-of-plane momentum $p$, so we introduce also $\Delta p$ to model this but always centered at $p_0=0$. For us, the case of $p_0=0$ is most relevant but this can be easily extended to any $p_0$. This leads to our final average loss function
\begin{equation}
L(\omega,q_0) = \frac{2}{\sqrt{\pi} \Delta p} \int_0^{\infty} dp e^{-\left(\frac{p}{\Delta p}\right)^2} L(\omega,q_0,p) . 
\end{equation}
We have plotted this in Fig. \ref{fig:lf3} and we see that for smaller uncertainty $\Delta p$ only the $1.0$ eV is visible, while for larger uncertainty in $p$ there are contributions from both small and large $p$. Such that there are two peaks in the loss function. This is not due to the fact that there are two modes, but due to the $p$ dependence of the plasmon mode. We especially display this for $q_0=0$, because this is a value of the in-plane momentum with a great dependence on $p$. For larger in-plane momenta the uncertainty in $p$ does not make a significant difference.

\section{Conclusions and Discussion}
In this paper we discussed a layered strange metal and computed the density-density response of this system. More specifically, we considered a cuprate with a bilayer crystal structure, as is the case for Bi-2212. We modeled the strong short-range interactions in each CuO$_2$ layer using the holographic Gubser-Rocha model, and we obtained the associated density-density response function for these strong interactions from the AdS/CFT correspondence. In addition, we incorporated the long-range Coulomb interactions by means of a double-trace deformation, which results in the density-density response function of the layered strange metal. We calculated the density-density spectral function for arbitrary values of the out-of-plane Bloch momentum $p$, and find both an in-phase and an out-of-phase mode. We computed the dispersion of these modes and showed that the in-phase plasmon mode is gapped for $p=0$, while it has an acoustic nature at long wavelengths for nonzero $p$. The out-of-phase mode always has an acoustic nature in the long-wavelength limit. Furthermore, we extracted the conductivity of the bilayer crystal from the density-density response function, taking into account the disorder that is present in experiments by introducing Planckian dissipation. In the parameter regime typical for cuprates there is always a Drude peak visible in the conductivity with a dc-resistivity linear in temperature.
Finally, we used the total density-density response function to construct the loss function which is measured in transmission EELS and we discuss its behavior.
In principle, the loss function only contains a single peak belonging to the in-phase plasmon mode, since the intensity of the out-of-phase mode is smaller in the relevant regime of $p$. 
Only when allowing for a large experimental uncertainty in the out-of-plane momentum $p$, and with a transverse momentum close to zero, there are two wide peaks visible. The mode around $1.0$ eV is due to the contributions close to $p=0$, while the lower energy peak arises from contributions with a larger value of $p$.

Throughout this paper, we made a number of assumptions to simplify the system.
For example, in the concrete example of Bi-2212 that we considered, there is a difference between the atomic structure in between the pair of layers close to each other and in between the pairs of layers. Therefore, the different dielectric constants might quantitatively influence the behavior of plasmons. Another assumption we made is that there is rotational invariance in each layer, which is of course not exactly the case, since there is a square lattice structure in the CuO$_2$ layers. 
Although this lattice structure does not influence the dispersion of the plasmon for small $q$, it does play a role for larger values of $q$. Hence, it would be interesting to include the lattice in the future.
Besides, many cuprates are known to have an extra periodicity which modulates their atomic lattice, known as the supermodulation \cite{tsuei2000,takeo2014}, 
which may lead to additional phenomena such as charge-density waves and Umklapp scattering \cite{chang2012}.

The findings in this paper provide new insights into the plasmons in layered strange metals. In particular, we notice that the holographic Gubser-Rocha model can reproduce the acoustic plasmon branches that have been observed in RIXS experiments on cuprates. However, our theoretical predictions appear to contradict the EELS results which are obtained by our experimental research group, in which no acoustic plasmon was observed \cite{thijs2023}. This apparent contradiction certainly challenges the experimentalist to either discover the right experimental conditions to observe the acoustic plasmon contribution or come up with arguments to explain why an acoustic plasmon cannot be measured. Or perhaps we should revise our theoretical framework and include potentially relevant features which it lacks at the moment, such as the lattice, phonons, charge-density waves, and so on. In any case, it is evident that these latter measurements require more extensive analysis. On top of that, we hope that the theory presented in this paper stimulates further experiments regarding plasmons in strange metals.

\section*{Acknowledgements}
This work is part of the D-ITP consortium, a program of the Dutch Research Council (NWO) that is funded by the Dutch Ministry of Education, Culture and Science (OCW).

\end{document}